\documentclass[useAMS,usenatbib]{mn2e}
\usepackage{graphicx,amsmath,multirow,amssymb}
\usepackage{natbib}

\newcommand\apj{{ApJ}}%
\newcommand\aap{{A\&A}}%
\newcommand\mnras{{MNRAS}}%
\newcommand\memsai{\ref@jnl{Mem.~Soc.~Astron.~Italiana}}%
\newcommand{\comment}[1]{}

\newcommand\ocen{$\omega$ Cen}
\def\simgt{\lower.5ex\hbox{$\; \buildrel > \over \sim \;$}}
\def\simlt{\lower.5ex\hbox{$\; \buildrel < \over \sim \;$}}

\title[Multiple populations in 47 Tuc]
{The formation of multiple populations in the globular cluster 47 Tuc}
\author[P.Ventura et al.]{P. Ventura$^{1}$\thanks{E-mail:
paolo.ventura@oa-roma.inaf.it (AVR)}, M. Di Criscienzo$^{2}$, F. D'Antona$^{1}$, E. Vesperini$^{5}$, 
M. Tailo$^{1,3}$, 
\newauthor
F. Dell'Agli$^{1,3}$, A. D'Ercole$^{4}$\\
$^{1}$INAF-Osservatorio Astronomico di Roma, Via Frascati 33, Monte Porzio Catone 00040, Italy\\
$^{2}$INAF-Osservatorio Astronomico di Capodimonte, Salita Moiariello 16, Napoli 80131, Italy\\
$^{3}$Dipartimento di Fisica, Universit\`a di Roma ``La Sapienza'', Italy\\
$^{4}$INAF-Osservatorio Astronomico di Bologna, Via Ranzani 1, Bologna 40127, Italy\\
$^{5}$Department of Astronomy, Indiana University, Bloomington, USA\\}
\begin{document}

\date{Accepted, Received; in original form }

\pagerange{\pageref{firstpage}--\pageref{lastpage}} \pubyear{2002}

\maketitle

\label{firstpage}

\begin{abstract}
We use the combination of photometric and spectroscopic data of 47 Tuc stars to reconstruct
the possible formation of a second generation of stars in the central regions
of the cluster, from matter ejected from massive Asymptotic Giant Branch stars, diluted 
with pristine gas.

The yields from massive
AGB stars with the appropriate
metallicity (Z=0.004, i.e. [Fe/H]$=-0.75$) are compatible with the observations, in terms 
of extension and slope of the patterns observed, involving oxygen, nitrogen, sodium and
aluminium.

Based on the constraints on the maximum 
helium of 47 Tuc stars provided by photometric investigations, and on the helium content
of the ejecta, we estimate that the gas out of which second generation stars formed was 
composed of about one-third of gas from intermediate mass stars, with M$\geq 5$M$_{\odot}$ 
and about two-thirds of pristine gas. We tentatively identify the few stars whose
Na, Al and O abundances resemble the undiluted AGB yields with the small fraction of
47 Tuc stars populating the faint subgiant branch.

From the relative fraction of first and second generation stars currently observed, we estimate
that the initial FG population in 47 Tuc was about 7.5 times more massive than the cluster 
current total mass.
\end{abstract}


\section{Introduction}
The traditional paradigm that stars in Globular Clusters (GC) are an example of a
coeval and chemically homogeneous population was challenged by photometric and spectroscopic
results, which highlighted the presence of at least two generations of stars: a first
generation (FG), with the chemistry of the cloud from which the cluster formed, and
an additional component (second generation, hereinafter SG), with a composition showing 
the signature of proton--capture nucleosynthesis.

The surface abundances of stars belonging to the SG define abundance patterns, where 
sodium is correlated to aluminium, and anticorrelated to magnesium and oxygen.
The extension and the slope of these trends change from cluster to cluster, but it
is interesting that they have been detected within each GC investigated so far 
\citep{gratton04, gratton12}. These chemical anomalies involve only ''light'' elements, 
up to silicon, whereas no spread is observed in the iron content\footnote{There are indeed 
a few clusters in which an iron variation is present, such as \ocen\ 
\citep[e.g.][]{norris1996} and M 22 
\citep[e.g.][]{marino09}. Their chemical evolution must have been more complex than in 
standard mono--metallic clusters.}.

On the photometric side, the hypothesis that the morphology of the Horizontal
Branch (HB) of some GCs could be explained by the presence of two or more
populations, differing in their original content of helium, came from the seminal paper
by \citet{dantona02}, and subsequently extended to NGC 2808 \citep{dantona04}, 
M3 and M13 \citep{caloi05}, NGC 6441 \citep{caloi07}.

These early speculations were later confirmed by detailed photometric analysis of the
Main Sequence (MS) of the cluster NGC 2808, which was shown to be split in three
components, differing in their helium content \citep{dantona05, piotto07}. Subsequent
investigations detected a splitting in the MS of NGC 6752 \citep{milone10}, and in the
subgiant branches (SGBs) of NGC 1851, NGC 6656, 47 Tuc, and other five GCs
\citep{milone08, marino09, anderson09, piotto12}.

This impressive set of evidence indicates that in most (if not all) GCs a self--enrichment
mechanism must have produced gas contaminated by p--capture nucleosynthesis, such that
one or more additional generations of stars formed, and are currently co-existing with the
original population.

According to the most complete scenario suggested so far, a crucial role as
polluters of the intra--cluster medium was played by stars of intermediate mass during
the Asymptotic Giant Branch (AGB) phase \citep{ventura01}. The paper by \citet {dercole08}  
set the theoretical framework to describe the formation of SG stars in GCs, by gas
ejected by AGBs possibly mixed with pristine gas, survived to the epoch of supernovae 
explosions. This approach allowed to reconstruct the formation of multiple populations
in  M4 (an example of a cluster showing mild anomalies) and NGC 2808 (example of a
cluster hosting, in addition, an extremely helium rich population) \citep{dercole10,dercole12}.

As we will show in this paper to reproduce the observed abundance patterns processed 
gas provided only by AGB stars with masses between about 5M$_{\odot}$ and 8M$_{\odot}$ can be 
used along with some pristine gas. As already discussed in several of the papers cited above, 
the amount of gas available for SG formation in this scenario (as well in other competing 
scenarios for which a detailed study of the resulting abundance patterns have been investigated; 
see e.g. \citet{bekki2007, dercole08, decressin2007, renzini2008, carretta09, bekki2011}) 
implies that the FG cluster must have initially been more massive. We discuss the specific 
implications of the model presented here in section 4. 

The globular cluster NGC 104, better known as 47 Tuc, is a valuable test to understand
the formation process of multiple populations in globular clusters. Until a few years
ago it was considered as the prototype of a single stellar population, based on the
photometric morphology of the MS and the HB. The first challenge to this belief came from
the analysis based on the HST archival data by \citet{anderson09}, showing the splitting
of the SGB in two components, separated in magnitude. 
\citet{marcella10} then showed that the morphology of the HB of 47 Tuc is consistent with the
presence of two populations, differing in helium content up to a maximum of
$\Delta Y \sim 0.03$. Such interpretation was recently confirmed by the detailed 
photometric analysis published in \citet{milone12}, who explained the complex of the observed 
colours of MS stars with a couple of populations, one with the primeval composition, and 
another with a small helium enhancement, which, in agreement with \citet{marcella10}, 
is within $\Delta Y=0.03$.

From a spectroscopic point of view, a bimodal distribution of CN band strengths among
giant stars belonging to 47 Tuc was found by \citet{briley97}. The following
analyses by \citet{cannon98} and \citet{harbeck03}, showed that
this bimodality still exists 2.5 mag below the Turn Off, suggesting the presence of
two populations in the cluster, differing in their chemistry. The presence of a O--Na 
anticorrelation in unevolved stars was first detected by 
\citet{carretta04}, based on the spectroscopic analysis of 7 objects, confirmed
by \citet{carretta09} and by \citet{dorazi10}, who investigated a much more complete 
sample, made of 109 sources. A step forward was made by \citet{gratton13}, who found a
correlation among the color of HB stars and the abundances of oxygen, sodium and
aluminium, and by \citet{carretta13}, who presented the chemical patterns
defined by 116 giants in 47 Tuc.

This rich set of data allows one to test whether the self--enrichment mechanism by AGBs
can explain the patterns traced by stars in 47 Tuc, and account for the distribution
of HB and MS stars in the Color--Magnitude diagram. For this purpose, we calculated a set
of AGB models having an input chemistry appropriate for stars in this cluster, we determined the
composition of the ejecta, and tested whether the observed patterns' extension and slope are
reproduced. We use work by \citet{marcella10} and \citet{milone12} to determine the dilution with pristine
material needed to allow the maximum spread of helium detected: this will set a limit
to the most extreme chemistry that can be obtained.

\section{The stellar models}
\label{models}

\subsection{Numerical and physical inputs}
The evolutionary sequences used in the present investigation were calculated by means
of the ATON code for stellar evolution. The interested reader may find in \citet{ventura98}
a detailed description of the code. The micro-- and macro--physics adopted are the same
as in the recent investigations by our group on this topic, and can be found. e.g., in 
\citet{vd09}. 

In regions unstable to convective motions, the temperature gradient is found via the 
Full Spectrum of Turbulence (hereianfter FST) model \citep{cm91}. Mixing of chemicals is 
coupled to nuclear burning via a diffusive approach, using the schematization by 
\citet{ce76}. The overshoot of convective bubbles into
radiatively stable regions is described by an exponential decay of convective velocities,
characterized by an e--folding distance proportional to the parameter $\zeta$. For all
the evolutionary phases preceding the AGB phase we chose $\zeta=0.02$, in agreement
with our previous explorations, based on the calibration given in \citet{ventura98}.
For the AGB evolution, we calibrated the possible extra--mixing from the base of the
convective envelope and from the borders of the convective shell which forms following 
the ignition of thermal pulses in order to reproduce the luminosity function of carbon
stars in the Large Magellanic Cloud. By following this approach, similar to \citet{marigo07}, 
we find $\zeta=0.002$.
 
The range of masses involved is $1M_{\odot} \leq M \leq 8M_{\odot}$. Models of smaller
mass hardly experience the AGB phase, and are however not of interest for the scope of this
paper. Stars of mass exceeding 8M$_{\odot}$ are expected to undergo core--collapse,
thus skipping the AGB phase. For each mass we followed the evolution from the pre--main
sequence up to the almost complete loss of the convective envelope. Models of mass below
2M$_{\odot}$ experience the helium flash at the tip of their red giant phase. The evolutionary
sequences for these masses were re--started from the horizontal branch, where 
He--burning takes place in the non degenerate core resulting from the flash.

Stars of mass above 6.5M$_{\odot}$ undergo carbon ignition in a 
degenerate layer above the core: these models develop a core made up of oxygen and neon,
and experience a series of thermal pulses, as their lower mass counterparts 
\citep{garcia97, ritossa96, ritossa99, siess06, siess07, siess10}.

\begin{table*}
\caption{Relevant properties of $Z=4\times 10^{-3}$ AGB and SAGB models}
\label{yields}
\begin{tabular}{cccccccccccc}
\hline
\hline
M$/$M$_{\odot}$ & $\tau_{\rm evol}$ & M$_c/$M$_{\odot}$ & T$_{\rm bce}^{\rm max}$ & Y & 
Li & [C/Fe] & [N/Fe] & [O/Fe] & [Na/Fe] & [Mg/Fe] & [Al/Fe]  \\
\hline
1.0   &  6.4e9   & 0.53   & 2.4e6    &  0.283 &  -1.66  &   0.62  &  0.04  &  0.22   &  0.02  &   0.20  &  0.00     \\
1.25  &  3.1e9   & 0.54   & 3.1e6    &  0.284 &  -1.93  &   0.81  &  0.03  &  0.23   &  0.04  &   0.20  &  0.00     \\
1.5   &  2.0e9   & 0.55   & 3.8e6    &  0.287 &  -1.90  &   1.04  &  0.04  &  0.26   &  0.08  &   0.20  &  0.00     \\
2.0   &  1.1e9   & 0.50   & 4.1e6    &  0.270 &   1.49  &   1.47  &  0.45  &  0.56   &  0.43  &   0.25  &  0.20     \\
2.5   &  5.7e8   & 0.57   & 5.5e6    &  0.296 &   1.40  &   1.63  &  0.54  &  0.76   &  0.64  &   0.47  &  0.73     \\
3.0   &  3.4e8   & 0.70   & 2.3e7    &  0.283 &   1.90  &   1.24  &  0.47  &  0.42   &  0.33  &   0.31  &  0.55     \\
3.5   &  2.3e8   & 0.79   & 8.1e7    &  0.283 &   2.80  &   0.04  &  1.37  &  0.19   &  1.01  &   0.22  &  0.20     \\
4.0   &  1.7e8   & 0.82   & 8.7e7    &  0.304 &   2.34  &  -0.11  &  1.37  &  0.08   &  0.93  &   0.21  &  0.27     \\
4.5   &  1.3e8   & 0.85   & 9.2e7    &  0.324 &   2.42  &  -0.20  &  1.30  & -0.06   &  0.97  &   0.20  &  0.40     \\
5.0   &  1.0e8   & 0.89   & 9.6e7    &  0.339 &   2.33  &  -0.24  &  1.23  & -0.13   &  0.70  &   0.18  &  0.49     \\
5.5   &  8.4e7   & 0.96   & 1.01e8   &  0.347 &   2.35  &  -1.00  &  1.01  & -0.18   &  0.57  &   0.17  &  0.52     \\
6.0   &  6.7e7   & 1.01   & 1.05e8   &  0.357 &   2.44  &  -1.04  &  1.08  & -0.19   &  0.54  &   0.17  &  0.50     \\
6.5   &  5.9e7   & 1.07   & 1.07e8   &  0.356 &   2.45  &  -1.06  &  1.05  & -0.18   &  0.52  &   0.16  &  0.48     \\
7.0   &  5.1e7   & 1.18   & 1.11e8   &  0.361 &   2.49  &  -1.06  &  1.03  & -0.15   &  0.53  &   0.16  &  0.46     \\
7.5   &  4.4e7   & 1.27   & 1.14e8   &  0.367 &   2.74  &  -1.07  &  1.02  & -0.13   &  0.54  &   0.17  &  0.44     \\
8.0   &  4.0e7   & 1.33   & 1.24e8   &  0.375 &   3.24  &  -1.05  &  1.01  & -0.12   &  0.57  &   0.17  &  0.43     \\
\hline
\end{tabular}
\end{table*}

The initial composition in terms of metal and helium abundance is Z=0.004 and Y=0.26.
This chemistry is similar to the models presented in \citet{ventura08a}, based on the
mass fractions of the individual species given by \citet{gs98}; the difference is
that in the present investigation the mixture is assumed to be alpha--enhanced, with
[$\alpha/$Fe]=+0.2 (in \citet{ventura08a} we used [$\alpha/$Fe]=+0.4). 
These choices,
assuming a solar metallicity Z$_{\odot}=0.017$, correspond to an iron content [Fe/H]$=-0.75$, 
the same measured in 47 Tuc stars \citep{carretta09}. In addition to \citet{ventura08a} paper 
we calculated the SAGB evolution of stars with mass M$>6.5$M$_{\odot}$.

\subsection{AGB evolution and stellar yields}
Following the pioneering investigations by \citet{schw65, schw67}, we know that
stars of intermediate mass, after the end of core helium burning, experience a
series of thermal pulses (TPs), in the evolutionary phase known as Asymptotic
Giant Branch. An updated, exhaustive description of the main features of the 
AGB evolution can be found, e.g., in \citet{herwig05} and \citet{karakas11}.

\begin{figure*}
\begin{minipage}{0.33\textwidth}
\resizebox{1.\hsize}{!}{\includegraphics{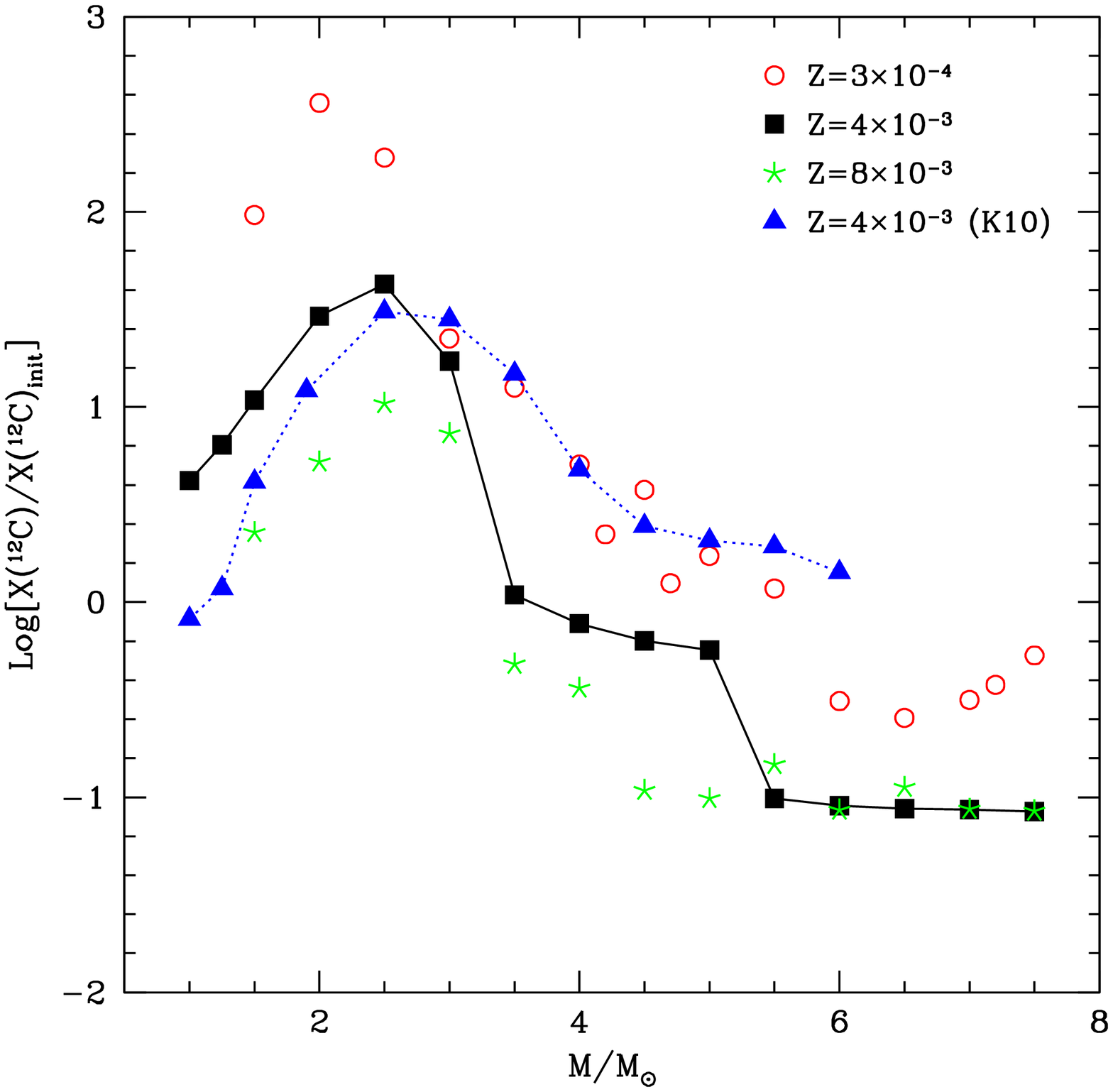}}
\end{minipage}
\begin{minipage}{0.33\textwidth}
\resizebox{1.\hsize}{!}{\includegraphics{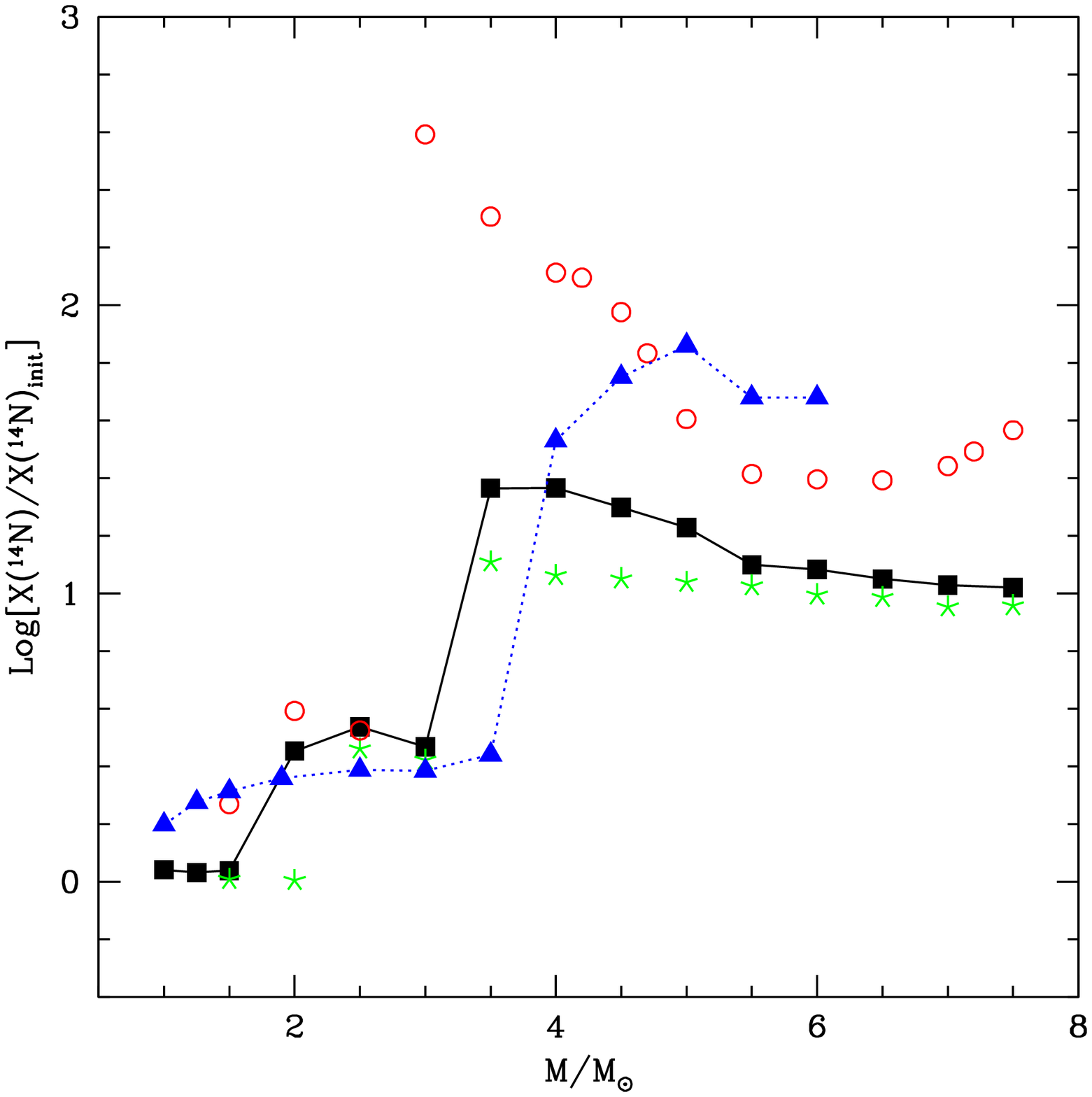}}
\end{minipage}
\begin{minipage}{0.33\textwidth}
\resizebox{1.\hsize}{!}{\includegraphics{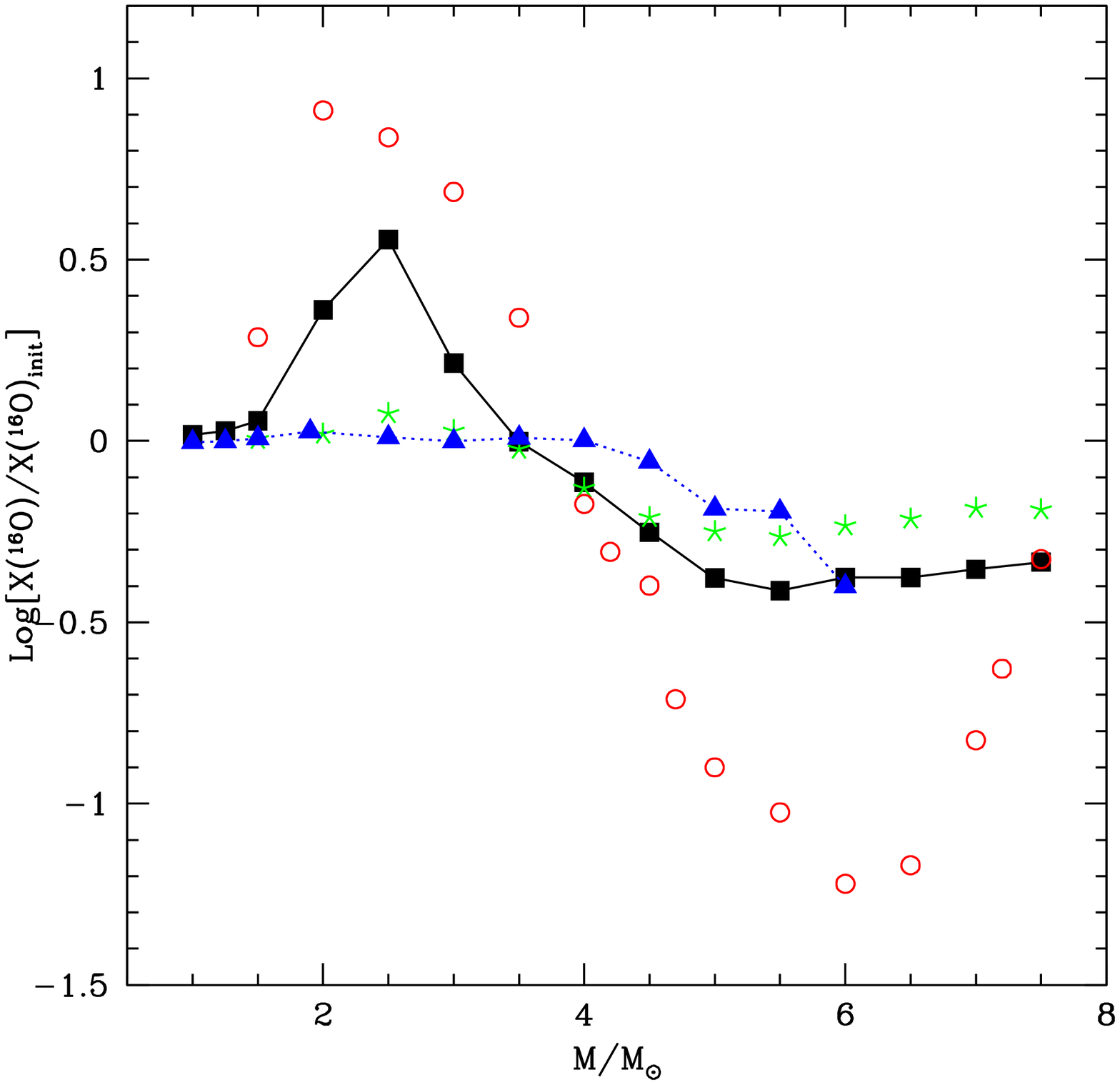}}
\end{minipage}
\caption{The average mass fractions of carbon (Left), Nitrogen (middle) and Oxygen (Right)
in the ejecta of AGB and SAGB stars, as a function of the initial mass. The ordinate
shows the ratio of the abundance of the elements to the initial mass fraction.
The results presented here, with metallicity Z=0.004, are indicated with black, full squares,
and are connected with solid lines. The blue triangles indicate the results by K10, and are
connected with a dotted curve. Red, open points and green asterisks show, respectively, results for
metallicities Z$=3\times 10^{-4}$ and Z$=8\times 10^{-3}$ by \citet{ventura13}.
}
\label{CNO}
\end{figure*}

The composition of the gas ejected by AGBs depends on the interface between the two
mechanisms able to alter the surface chemistry: a) Third Dredge--Up (TDU), i.e. the inwards
penetration of the convective envelope down to layers previously touched by 3$\alpha$
activity, and b) Hot Bottom Burning (HBB), with the activation of p--capture nucleosynthesis
at the bottom of the surface convective layer \citep{renzini81, blocker91}. 
TDU favours the increase of the surface carbon, and is the dominant mechanism in low--mass 
AGBs, with M$\leq 3$M$_{\odot}$. More massive stars may experience HBB, provided that the 
temperature at the base of their envelope exceeds $\sim 40$MK.

The efficiency of both mechanisms is unfortunately sensitive to the macro--physics
description used. The extent of TDU depends on the treatment of convective borders, 
particularly to the extra--mixing from the base of the convective envelope and 
the boundaries of the convective shell which forms after each TP.

The strength of HBB is extremely sensitive to the convective model adopted
\citep{ventura05}. Models calculated with the Full Spectrum of Turbulence
description of the convective regions are found to experience a much stronger HBB
in comparisons with models of same mass and metallicity calculated by means of the
traditional Mixing Length Theory of turbulent convection. 

Table \ref{yields} summarizes the main physical and chemical results for the models
investigated. For each mass we show the evolutionary time scale, the core mass at the
beginning of the AGB phase, the maximum temperature achieved at the bottom of the convective
envelope, and the average composition of the ejecta. For the i--th element we indicate the quantity
[i/Fe]=$\log$(X$_i$/X(Fe))-$\log$(X$_i$/X(Fe))$_{\odot}$. For helium and lithium we show, respectively,
the average mass fraction in the ejecta (Y), and the quantity 
$\log$($\epsilon$(Li))=$\log$(n(Li)/n(H))+12.

Col.~5 of Table 1 shows that the ejecta of AGBs are helium--rich, as a consequence of the
two dredge--up episodes following the core H-- and He--burning phases, and, for the models
experiencing HBB, of the p--capture nucleosynthesis at the bottom of the surface convective
layer. Models with mass exceeding 5M$_{\odot}$ produce ejecta with a helium mass
fraction $Y \sim 0.35$, almost 0.1 larger than the initial abundance. This prediction,
unlike those regarding the other species, is rather robust, because most of the helium
enrichment occurs during the second dredge--up, which makes this finding independent
of the uncertainties affecting the following thermal pulses phase \citep{ventura09}.

\begin{figure}
\begin{minipage}{0.50\textwidth}
\resizebox{1.\hsize}{!}{\includegraphics{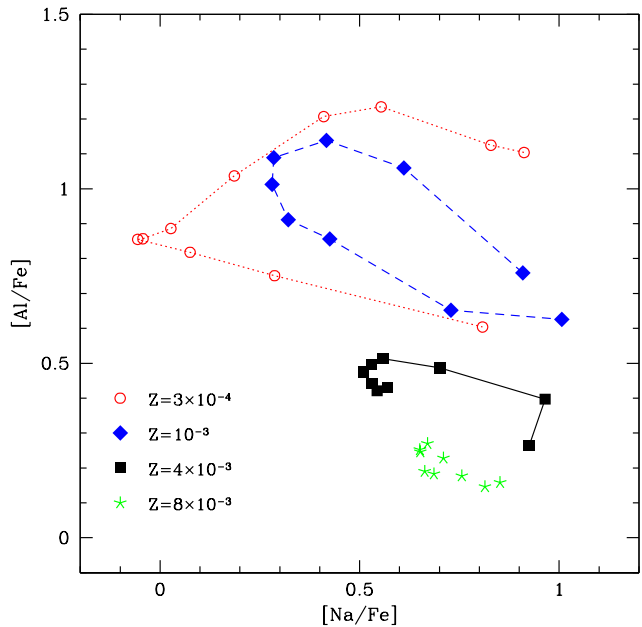}}
\end{minipage}
\caption{Sodium and Aluminium abundances of the gas expelled by AGB stars in the
[Na/Fe]--[Al/Fe] plane. The results presented in this work, connected with a solid line, 
are indicated with full, black squares. The other points indicate
results for different metallicities published in \citet{ventura13}. 
For the sake of clarity, we only show
yields for masses experiencing HBB, with M$\geq 4$M$_{\odot}$. For each metallicity,
results for increasing mass are in the counter-clockwise direction.
}
\label{naal}
\end{figure}

The three panels of Fig.~\ref{CNO} show the average content of the gas ejected as a 
function of the initial mass, in terms of the mass fractions of carbon (left panel),
nitrogen (middle) and oxygen (right). To better understand the extent of the variation
from the initial content of each species, we show the ratio between the average and
the initial abundance. The present results are compared with previous findings 
for different metallicities published in \citet{ventura13}, and with models of the same
metallicity by \citet{karakas10} (hereinafter K10).

Both the present models and those by K10 show an increase in the carbon content of the
ejecta in the low--mass domain, for M$\leq 3$M$_{\odot}$: this is due to the repeated TDU 
episodes, that transport to the stellar surface carbon synthesized in the $3\alpha$ burning 
shell. The trend of X($^{12}$C) with mass is rather similar in the two cases, as also the
maximum increase in the carbon content, found to be $\delta$(C)$\sim 1.5-1.6$ dex.
The results between this investigation and the work by K10 are different for 
$M \geq 3.5$M$_{\odot}$, because our models experience a stronger HBB, with a faster 
destruction of the surface carbon: while the yields by K10 are enriched in 
carbon for all masses, our models experiencing HBB show a carbon decrease, 
which reaches the asymptotic value of $\delta$(C)$\sim -1$ dex in the SAGB domain.

The different extent of the HBB experienced is also the reason for the difference in the
oxygen content of the ejecta (see right panel) for $M \geq 3.5$M$_{\odot}$: in the 
present compilation the oxygen is systematically lower than in K10, with the only exception of 
the 6M$_{\odot}$ model, which shows the same depletion of $\delta$(O)$\sim -0.4$ dex
in the two cases.

The trend of the nitrogen abundance with mass shows an abrupt increase
around $\sim 3$M$_{\odot}$, because in that range of mass the carbon dredged--up to
the surface is converted into nitrogen during the following interpulse phase. Note that
in the $M \geq 3.5$M$_{\odot}$ domain our nitrogen yields are lower than K10, because the 
stronger HBB experienced by our models limits the number of thermal pulses, and thus the 
amount of carbon available.

An important consequence of the strong HBB in the envelope of massive AGB
models calculated with the FST description of convection is that TDU is scarcely
efficient in modifying the surface chemistry: as discussed previously, the star
loses the whole envelope after a limited number of TPs, before TDU becomes
efficient \citep{ventura08b}. This, in turn, implies that the overall C+N+O
content of the ejecta of massive AGBs (with M$> 5$M$_{\odot}$) and SAGBs is practically
unchanged with respect to the initial abundance, as expected from a pure p--capture
nucleosynthesis \citep{vd09}.

\begin{figure*}
\begin{minipage}{0.45\textwidth}
\resizebox{1.\hsize}{!}{\includegraphics{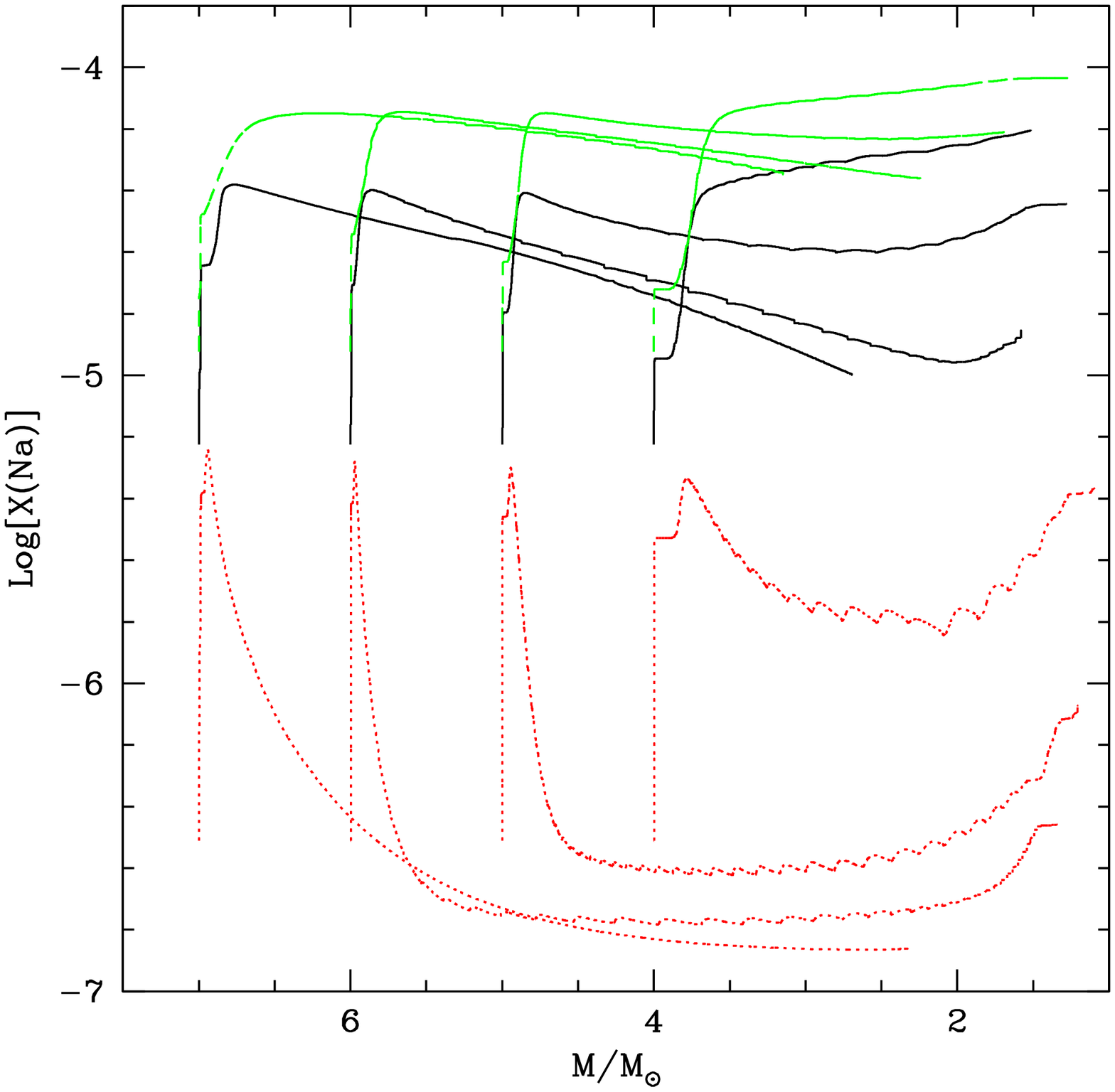}}
\end{minipage}
\begin{minipage}{0.45\textwidth}
\resizebox{1.\hsize}{!}{\includegraphics{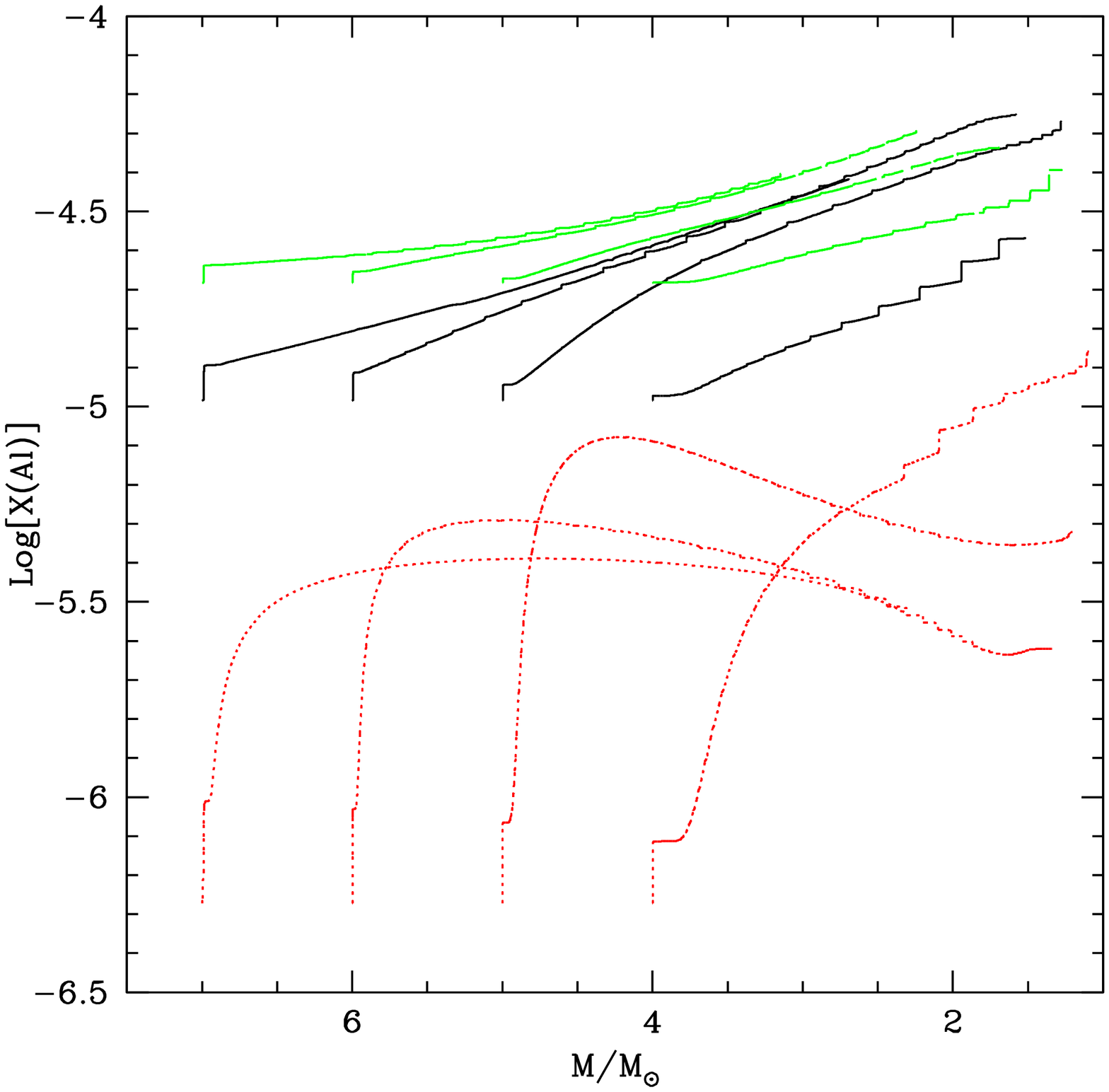}}
\end{minipage}
\caption{The variation during the AGB phase of the surface abundances
of sodium (Left) and Aluminium (Right) in AGB models of initial 
mass 4,5,6,7M$_{\odot}$ and various metallicities. The total mass of the star 
(descreasing during the evolution) is reported on the abscissa. The models
discussed in this work are indicated with full, black lines. Red, dotted curves indicate
the results for Z$=3\times 10^{-4}$, whereas green, dashed lines refer to
Z$=8\times 10^{-3}$ models.
}
\label{yield}
\end{figure*}

The differences between models with different chemistry are due to 
the different impact of TDU and HBB in models with different metallicities.
In the low mass domain, where the yields are dominated by TDU, the percentage increase 
in the surface carbon is larger for models with lower Z, because the same amount
of carbon transported to the surface produces a larger increase in the surface
abundance. The lower metallicity models experience a stronger HBB \citep{ventura13}: this 
is the reason why models with Z=$3\times 10^{-4}$ have a much smaller oxygen content than 
their higher Z counterparts. The opposite behaviour is found in the 
Z=$8\times 10^{-3}$ case, where only a modest depletion of the surface oxygen is achieved.

We note in the right panel of Fig.~\ref{CNO} that in the range of massive AGBs and SAGBs
the low--Z models are characterized by a broad range of oxygen, while the models
with Z=$4\times 10^{-3}$ (and even more those with Z=$8\times 10^{-3}$) have a much smaller 
variation of oxygen
with mass. The reason for this difference is, again, the stronger HBB experienced by
massive AGBs of smaller metallicity: the fast increase in the surface luminosity favours
a large increase in the mass loss rate, such that in the models with the
largest core masses the envelope is lost rapidly, before an advanced nucleosynthesis 
can be achieved. The trend of the oxygen content of the ejecta with mass is consequently
not monotonic, the SAGB models showing up traces of a milder nuclear processing 
\citep{ventura11}.

In Fig.~\ref{naal} we show the sodium and aluminium content of the ejecta of the
models presented here, compared to models of different metallicities published in
\citet{ventura13}. In the [Na/Fe]--[Al/Fe] plane we only show the yields of masses 
experiencing HBB, with M$\geq 4$M$_{\odot}$.

The interpretation of the aluminium abundances is straightforward. Lower metallicity
models undergo stronger HBB, thus the Mg--Al nucleosynthesis proceeds faster.
This can be clearly seen in the right panel of Fig.~\ref{yield}, showing
the variation of the surface Al abundance in models of various mass and metallicity.
The increase of Al with respect to the initial abundance is enhanced in low--Z models,
whereas it is extremely small in the $Z=8\times 10^{-3}$ case. In the 
$Z=3\times 10^{-4}$ models (and also in the $Z=10^{-3}$ ones, not shown here for clarity 
reasons) the temperatures at the bottom of the surface convective zone become so large 
that eventually an equilibrium stage is reached, such that the rates of production and
destruction of Aluminium balance each other \citep{vcd11}: this is 
the motivation of the counter-clockwise shape
of the [Na/Fe]--[Al/Fe] trends in Fig.~\ref{naal} for these two metallicities, 
with the largest masses showing a smaller Al--enhancement in their ejecta.

In the $Z=4\times 10^{-3}$ models presented here the temperature at the bottom of the
convective envelope hardly exceeds $10^8$ K (see col.~4 of Table \ref{yields}), thus the 
Al--burning channel is never activated: the behaviour of the various masses involved is 
much more homogeneous (see right panel of Fig.~\ref{yield}), and the extent of the increase in 
Al in the ejecta is approximately independent of mass, for M$\geq 5$M$_{\odot}$: 
for the present mixture, massive AGBs produce winds with an average increase
in the Aluminium content of $\delta$[Al/Fe]$\sim +0.5$ dex.

The behaviour of sodium is more tricky, given the different sensitivity to temperature of 
the production and destruction channels \citep{ventura06}. While for temperatures below 
$\sim 70$MK sodium is produced at the expense of neon, for larger T's the destruction
process takes over. The typical behaviour of the surface sodium during the AGB evolution 
(see left panel of Fig.~\ref{yield}) shows an initial increase, followed by a depletion 
of the surface abundance in the latest evolutionary phases.

The higher is the temperature, the strongest is the rate at which sodium is consumed: 
lower metallicity models show on the average a lower content of sodium 
in the ejecta, as shown in Fig.~\ref{naal}. In low Z, SAGB models, mass loss is so fast to 
prevent a great destruction of the surface sodium \citep{ventura11}: this is the reason for the
large sodium content in the ejecta of the most massive models of $Z=3\times 10^{-4}$
and $Z=10^{-3}$. 

\begin{figure*}
\begin{minipage}{0.45\textwidth}
\resizebox{1.\hsize}{!}{\includegraphics{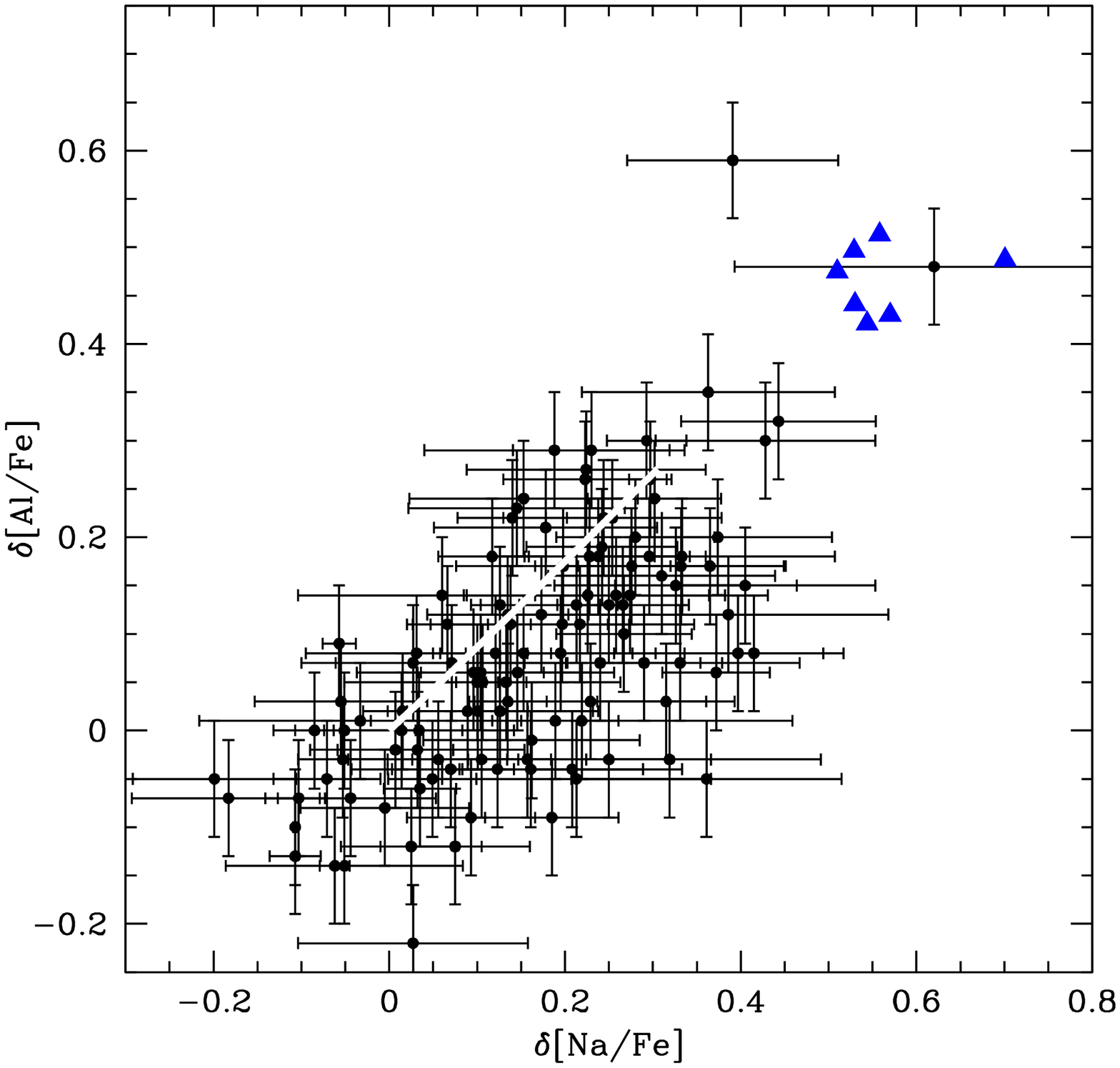}}
\end{minipage}
\begin{minipage}{0.45\textwidth}
\resizebox{1.\hsize}{!}{\includegraphics{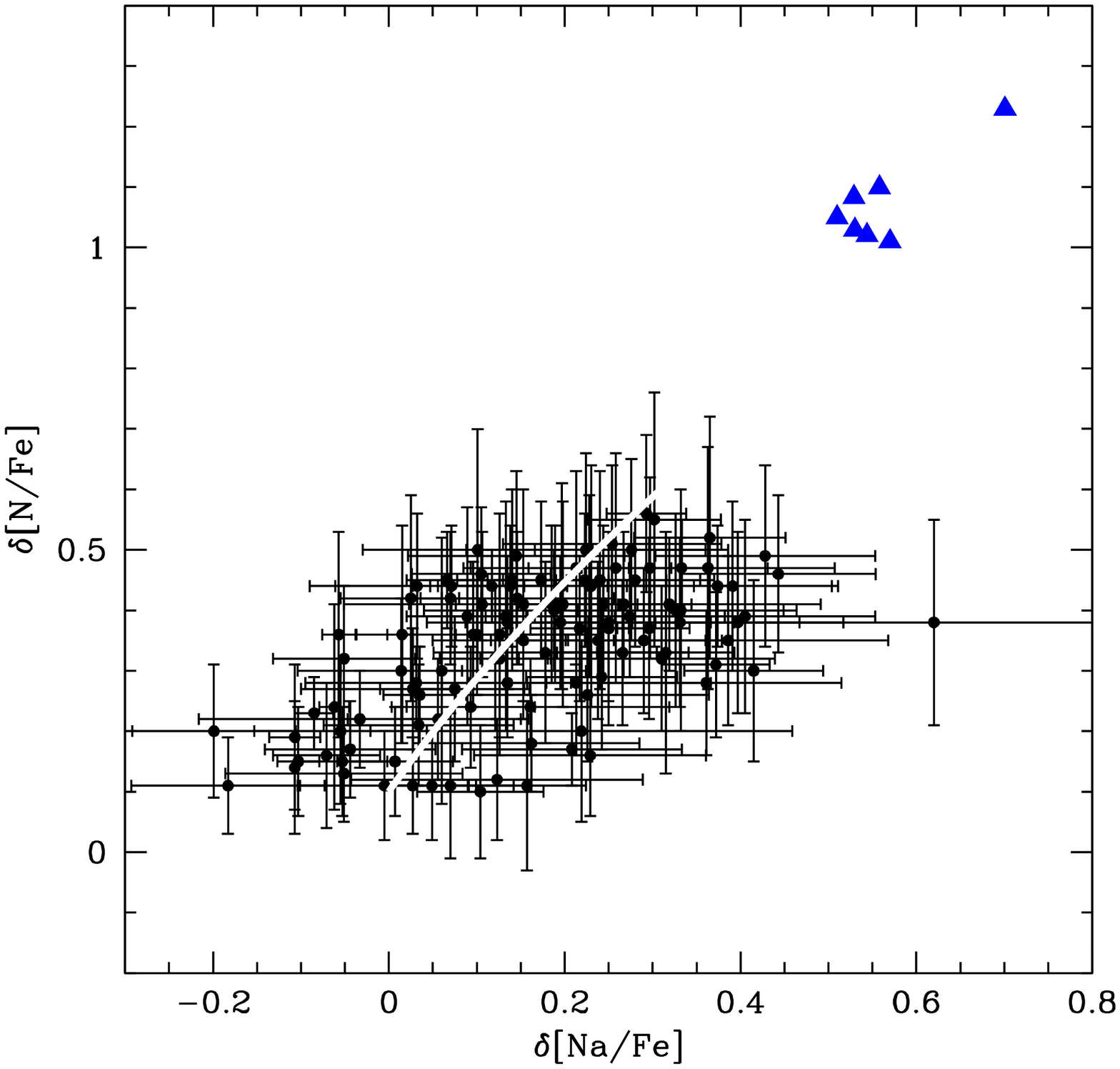}}
\end{minipage}
\caption{Left: Sodium and aluminium abundances observed in 47 Tuc stars by \citet{carretta13}
(black points) and in the ejecta of massive AGB and SAGB stars presented in Table \ref{yields}
(blue, full triangles). Only models with M$\geq 5$M$_{\odot}$ are shown. The observed 
abundances of sodium and aluminium were shifted, respectively, by -0.3 and -0.45 dex, 
to account for the offset between the initial abundances used in the models and the 
values observed in FG stars.
The white line is a dilution curve, obtained by mixing different fractions of unprocessed 
material and gas from AGBs (see text for details). Right: same as left panel,
but refereed to the Na--N plane. The shift applied to the observed nitrogen abundances is
-0.7 dex.
}
\label{carretta}
\end{figure*}

At $Z=4\times 10^{-3}$ sodium is initially produced in massive AGBs, the surface abundance
increasing by almost a factor $\sim 10$ compared to the initial mass fraction. Unlike
their lower Z counterparts, the destruction process proceeds later at a very slow pace. 
The sodium content of the ejecta is not very sensitive to the initial mass: we find
that the sodium increase in the gas lost is $\delta$[Na/Fe]$\sim +0.5$ dex.
 
To summarize our findings, we conclude that massive AGBs with the chemistry
examined here suffer a mild HBB. The gas ejected by these stars is expected
to present the signature of p--capture burning, with the depletion of the surface content
of oxygen of $-0.4$ dex, and an enhancement of the surface sodium and aluminium
of $+0.5$ dex. The nucleosynthesis experienced is not sufficiently advanced to
produce any modification of the surface silicon, whereas the overall content of
magnesium is poorly reduced with respect to the initial abundance, of 
$\sim 0.04$ dex\footnote{The overall magnesium content is only modestly touched 
by the nucleosynthesis experienced. However, the internal distribution among the
Mg isotopes is considerably different in comparison with the initial mixture:
most of the $^{24}$Mg is lost, whereas $^{25}$Mg is increased by almost one order
of magnitude.}. The gas ejected is helium--rich, with an
average helium of $Y \sim 0.35$. Finally, for what concerns the overall content of
CNO, the yields from models with mass above $5M_{\odot}$ are found to maintain the
initial C+N+O, whereas for $4$M$_{\odot} \leq$M$\leq 5$M$_{\odot}$ the increase in the
total CNO ranges from 20$\%$ up to almost $\sim 100\%$.

\section{The interpretation of the chemical patterns traced by 47 Tuc stars}

The recent investigation by \citet{carretta13} shows that stars in 47 Tuc trace well defined
abundance patterns, involving some of the elements touched by p--capture nucleosynthesis.
Sodium is correlated to aluminium and to nitrogen, and anticorrelated to oxygen. The present 
data do not allow to confirm possible variations in the abundances of magnesium and silicon,
but the variation of these elements, if present, cannot exceed a few percent. 

In the O--Na plane similar trends were found by the analysis focused on HB stars by
\citet{gratton13}, and by spectroscopy of unevolved stars of the same cluster \citep{dorazi10}.

To understand whether this set of observations can be explained on the basis of the 
self--enrichment mechanism by AGBs, we compare the data available with the chemistry 
obtained by mixing gas from massive AGBs and SAGBs with gas assumed to share the
primeval chemistry of the cluster, characterizing the FG component.

The dilution of the chemistry reported in Table \ref{yields} for M$\geq 5$M$_{\odot}$
with pristine material defines an abundance pattern, for different degrees of dilution.
Theoretically, all the results obtained with degrees of dilution ranging from 0 to 1 should be
considered: however, the photometric analysis of the HB \citep{marcella10} and of the MS of
47 Tuc \citep{milone12} indicate that the range of helium abundances of the stars in the
cluster is $\Delta$Y$\sim 0.03$, which rules out the possibility that uncontaminated stars
with the pure chemistry of AGBs formed: based on the values shown in Table \ref{yields}, these
stars should have a helium mass fraction Y$\sim 0.35$, which would determine a much wider
spread of the MS. These arguments allow to determine the minimum degree of dilution from
which the stars in the SG formed: mixing of pristine matter with Y$=0.26$ and gas from AGBs
with Y$=0.35$ leads to the maximum helium allowed from the photometric analysis, Y$=0.29$,
if we assume that the fraction of gas from AGBs used to form SG stars is one third, the
remaining being provided by pristine gas in the cluster. In the following analysis we will 
allow the relative contribution from gas ejected by AGBs to range 
from a minimum of 0$\%$ to a maximum of 35$\%$. Obviously stars formed with no AGB ejecta 
or with a very large fraction of pristine gas will have the chemical properties of FG stars 
and would not be classified as SG stars \citep{dercole11}.

Fig.~\ref{carretta} shows the results from \citet{carretta13} on the Na--Al (left panel) 
and Na--N (right) planes. In both axis we report the variation with respect to the abundances
observed in the stars that we assume to belong to the first generation of the cluster.
This choice is motivated by the offset between the solar--scaled abundances of nitrogen,
aluminium and sodium used in our models, and the minimum values for the same elements
reported in \citet{carretta13}. 

The AGB and SAGB yields reported in Table \ref{yields} are 
indicated with full, blue triangles. The white curve indicates a dilution relationship, 
obtained by mixing gas with the original chemistry of the cluster with matter ejected
by AGBs; this analysis is particularly simple in this case owing to the little variation 
with mass of the chemistry of the ejecta in the massive AGBs domain, and would be harder 
to be applied to lower metallicity clusters, for which the composition of the ejecta is 
much more sensitive to the initial mass of the star. The dilution curves in Fig.~\ref{carretta}
are limited in such a way that the contribution of AGB ejecta range from $0\%$ to a
maximum of $35\%$. As pointed out above, and previously discussed in \citet{dercole11}, 
stars forming from a mix of gas with a large fraction of pristine gas would fall in the 
FG portion of the chemical patterns planes. We extend the lines to such a large 
fraction of pristine gas only for illustrative purposes.

We see from both panels that the observed trends, within the error bars associated to the 
abundances of the individual stars, are satisfactorily reproduced. In the Na--Al
plane (left panel) most of the stars fall within the dilution curve, with the exception of 
two sources, that show a chemistry similar to the pure AGB ejecta. The same holds
for the Na--N pattern, the dilution curve encompassing the most extreme, nitrogen--rich
stars.

\begin{figure}
\begin{minipage}{0.52\textwidth}
\resizebox{1.\hsize}{!}{\includegraphics{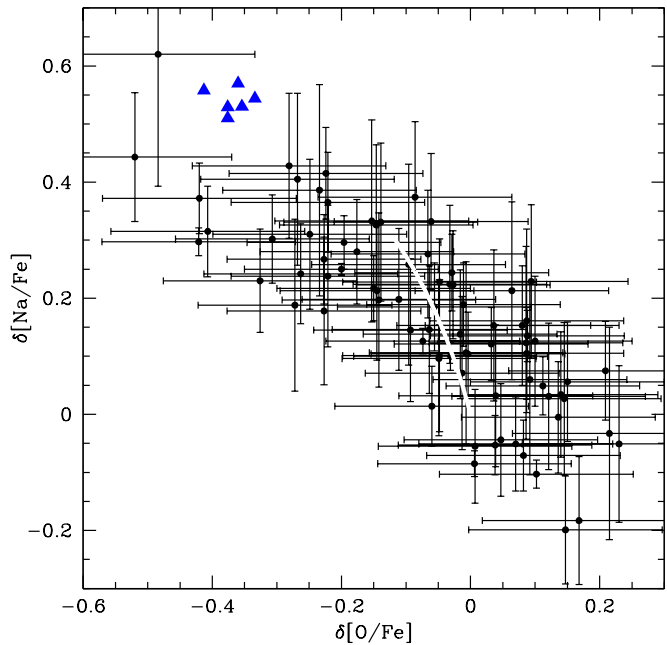}}
\end{minipage}
\caption{Same as Fig.~\ref{carretta}, but referring to the O--Na plane.
The sodium abundances observed were shifted as in Fig.~\ref{carretta}. Because the
initial oxygen in the models is [O/Fe]=+0.2, we applied a shift of -0.2 dex to the
yields. The observed oxygen abundances were also shifted by -0.2 dex, to account
for the difference between the initial oxygen in the models and the abundances
observed in FG stars (see text for details).}
\label{onacarretta}
\end{figure}

Fig.~\ref{onacarretta} shows the comparison between the theoretical predictions and the
observations, for what concerns the oxygen--sodium anticorrelation. Similarly to the
Na--Al and Na--N patterns showed in Fig.~\ref{carretta}, we show the (negative) variation
of the surface oxygen of the individual stars, assuming that the FG has an initial
oxygen [O/Fe]=+0.2.

The comparison in this
case is not straightforward, given the large uncertainties associated in particular to the
abundances of oxygen. The abundances of the ejecta lay in the upper--left portion of the plane,
where the most contaminated objects are found. Given the arguments discussed previously,
regarding the maximum degree of contamination by AGB gas in the formation of the second
generation of stars, we must rule out such extreme chemistries, limiting the contribution
from AGBs to $35\%$. In this way we obtain the dilution curve in Fig.~\ref{onacarretta}.
The observed points are reproduced within the error bars associated to the observations, 
with the exception of the three most extreme objects with the smallest abundances of
oxygen. This comparison, though less meaningful than the previous analysis based on the
abundances of aluminium and nitrogen, indicates that the HBB experienced by AGB models of
the same chemistry as 47 Tuc stars is appropriate to provide the oxygen depletion 
needed to fit the observations.

A word of caution concerning the assumptions made on the
chemical mixture from which the cluster formed (i.e. the composition of FG stars) is needed here.
The results from the above analysis hold provided that the values used
to build the models are correct, and that the observed data show an offset in the
measured abundances. Should the initial aluminium be that observed in the assumed FG 
component (a factor of $\sim 3$ higher than our assumption), the magnesium nucleosynthesis 
would work with the same efficiency, thus the amount al Aluminium produced, $\delta$(Al)=
$X(Al)_{\rm ejecta}-X(Al)_{\rm initial}$ would be unchanged: this would imply a smaller
percentage increase in the content of Al in the ejecta, thus a smaller difference 
between the abundances of SG and FG stars. This can be easily compensated by a higher 
content of magnesium, which we assumed to be [Mg/Fe]=+0.2, whereas data from
\citet{carretta13} point in favour of a larger abundance of [Mg/Fe]=+0.4.

The situation is less critical for sodium, not only because the offset between our
assumptions and the abundances given in \citet{carretta13} is limited to a factor $\sim 2$,
but also because the percentage increase in the surface sodium is only marginally
touched by the initial abundance: the horizontal extension of the dilution curves
in both panels of Fig.\ref{carretta} would be slightly shorter, but the conclusion
drawn would remain unchanged.

As far as oxygen is concerned, the rate of destruction of this species scales with its
abundance, which makes the percentage reduction independent on the assumed initial
value.

We do not discuss here the effects of a possible offset in the nitrogen content,
because the absolute values are strongly interfaced with the assumptions concerning the
initial carbon in the mixture, thus contain a high degree of arbitrariness. We limit
our analysis on the variations observed.

According to the models used here, no change in the silicon abundance should be detected.
This seems to be in agreement with the data from \citet{carretta13}, where no clear
Si--Al trend is observed. 

Magnesium is only marginally touched by the HBB experienced, the decrease in the overall
Mg being below 0.05 dex (see col.~ 11 of Table \ref{yields}). The detection of a similar spread 
among 47 Tuc stars is behind the possibilities of the present observations, given the 
undetermination in the measured abundances, exceeding 0.1 dex \citep{carretta13}.

\subsection{The faint turnoff in 47 Tuc}
The analysis by \citet{marcella10} suggested the presence in 47 Tuc of a stellar component
enriched in the overall C+N+O, based on the morphology of the SGB of the cluster. 
These faint turnoff stars should be revealed along the RGB. We suggest that they are the few stars
in the left panel of Fig. \ref{carretta} which are out of the dilution pattern but have
high [Na/Fe] and [Al/Fe]. The stars for which we have measures of oxygen have also low
[O/Fe], closer to the pure yields of our models.

Following the models by \citet{dercole12}, we can think that the bulk of SG stars in 47 Tuc
has been formed by dilution of pristine gas with the AGB ejecta. When the pristine gas gets 
consumed, the star formation may go on for a while, until type Ia supernovae completely
clean the cluster from gas. The stars formed in this phase would have the chemistry of the
5M$_{\odot}$ ejecta, scarcely diluted with pristine gas. The CNO enhancement of the 5M$_{\odot}$
is 1.35 times the original CNO, and the turnoff location of these stars could resemble 
the faint turnoff. This small population should also be helium rich, so their HB location 
would be among the most brilliant HB stars of 47 Tuc.

\citet{gratton13} show indeed that a couple of their bright star groups of 47 Tuc HB also
have high [Na/Fe] and low [O/Fe]. The rest of their bright sample consists of normal
FG stars, probably in an evolved phase out of the HB.

\section{47 Tuc: how the multiple populations formed}
The analysis carried out in the previous section shows that dilution of massive AGBs
ejecta with gas pristine allows to reproduce the observed patterns of 47 Tuc stars.
This is in agreement with the scenario described by \citet{dercole08} and 
\citet{dercole10, dercole11, dercole12} according to which
SG stars formed from a mix of AGB ejecta and pristine gas driven into the cluster central 
regions by a cooling flow and the SG formation process is halted after $\sim 100$ Myr 
by SN Ia explosions.

The chemistry with which SG stars form is therefore obtained by mixing processed matter 
with the chemical properties of the AGB and SAGB ejecta of stars evolving within 
$\sim 100$ Myr (i.e.
for mass M$\geq 5$M$_{\odot}$, see col.~2 of Table \ref{yields}) and pristine gas,
with the same composition of FG stars. The presence of pristine gas is an essential 
ingredient in the self--enrichment mechanism by AGBs (as well as in all the other models 
proposed in the literature; see D'Ercole 2011 for a discussion), and allows to reproduce 
the O--Na anticorrelation, otherwise inhibited by the direct correlation between the oxygen 
and the sodium yields by massive AGBs \citep{dercole11}.

In the case of 47 Tuc, our models suggest that to reproduce the photometric and spectroscopic 
observations SG stars must have formed out of a mix of gas in which about one third came from the
AGB and SAGB ejecta and the rest from pristine gas. 
According to \citet{carretta13} FG stars are $\sim 25\%$ of the
total population, the remaining $75\%$ being contaminated objects belonging to the SG.
The fact that the gas ejected by massive AGBs, for any realistic mass function, is only
$\sim 5\%$ of the mass of the initial FG population implies that a significant fraction 
of FG stars must have been lost by the cluster.
According to the simulations presented in  \citet{dercole08}, this early loss of FG stars 
(and the consequent increase in the fraction of SG stars) would occur during the cluster 
early evolution. Memory of the initial central concentration of the SG population would 
be preserved during this phase and would still be present in many clusters today 
\citep{vesperini13}. Based on the analysis presented in \citet{vesperini13}, 47 Tuc 
would be one of the clusters for which the SG should still be more concentrated in 
the inner regions than the FG population in agreement with what found in observational 
studies \citep{milone12,richer13}.

Following the calculation presented in \citet{dantona13} and applying it to the chemical 
model presented here in which SG stars formed from a mix composed AGB ejecta (one third) 
and pristine gas (two thirds), we estimate that the total initial FG mass of 47 Tuc was 
about 7.5 times the current cluster mass (assuming SG stars form with masses up to about 
8M$_{\odot}$ so that there are no SG SNII explosions \citep{dantona13}; 
the initial mass required would be smaller--about 4.5 times the current cluster mass--if 
one assumes a range of SG stars limited to 0.8 M$_{\odot}$).

\section{Conclusions}
In this paper we have studied the viability of a model to explain the abundance patterns 
defined by stars in 47 Tuc on the basis of the self--enrichment scenario by
massive AGBs.

For this purpose, We have specifically calculated a full set of AGB and SAGB models with the
chemistry of 47 Tuc stars, and iron content [Fe/H]$=-0.75$.

AGB models in the high mass (M$\geq 5$M$_{\odot}$) domain experience a soft Hot
Bottom Burning, producing ejecta enriched in Aluminium and sodium by +0.5 dex,
and depleted in their oxygen content by $-0.4$ dex. No meaningful magnesium and
silicon variations are expected. Similarly to other metallicities, the matter ejected 
is helium--rich, Y$\sim 0.35$.

Mixing of gas ejected by AGBs with pristine gas sharing the same chemistry as the 
stars originally present in the cluster allows to trace abundance patterns,
where sodium is correlated to nitrogen and aluminium, and anticorrelated to oxygen.
Based on the maximum helium enhancement allowed by the photometric analysis of the
HB and the MS of the same cluster, we estimate that SG stars formed from a mix of gas 
composed for about one third of AGB ejecta and two thirds of pristine gas. Using this 
constrain, we can fit the extension and the slopes of the various trends observed, 
provided that only stars with mass M$\geq 5$M$_{\odot}$ contributed to the formation of 
the second generation. This implies that the SG population formed formed within $\sim 100$ 
Myr, before the SN Ia explosions prevented further star formation.

Based on this analysis, and from the observed fraction of SG stars, $\sim 75\%$ of the total
stellar population, we estimate that 47 Tuc was initially about 7.5 times more massive than 
now (or 4.5 times for SG star formation limited to 0.8 M$_{\odot}$)
and that a large fraction of its initial FG population must have been lost during the cluster 
early evolution.

\section*{Acknowledgments}
Funding is acknowledged PRIN INAF 2011 “Multiple populations in GCs: their role in the 
Galaxy assembly” (PI: E. Carretta), and PRIN MIUR 2010–2011 “The Chemical and Dynamical 
Evolution of the Milky Way and Local Group Galaxies” (PI: F. Matteucci), prot. 2010LY5N2T. 
EV was supported in part by grant NASA-NNX13AF45G. MDC has been supported by the INAF 
fellowship 2010 grant.

\end{document}